\documentclass[12pt]{iopart}

\usepackage[utf8]{inputenc}
\usepackage[english]{babel}
\usepackage[T1]{fontenc}
\usepackage{graphicx}
\usepackage{bm}
\usepackage{xcolor} 
\usepackage{siunitx}
\usepackage[numbers,sort&compress]{natbib}

\newcommand{\mJ}{\milli\joule} 
 
\DeclareSIUnit\pxl{pixels}
\DeclareSIUnit\ppm{ppm}
\DeclareSIUnit\ppb{ppb}
\DeclareSIUnit\ppt{ppt}
\usepackage{cleveref}
\crefname{figure}{Figure}{Figures}
\crefname{equation}{Equation}{Equations}
\crefname{table}{Table}{Tables}

\graphicspath{{figures/}}

\newcommand*{\newblock}{}

\begin{document}

\title[HV Discharge acceleration by sequences of laser pulses]{HV discharge acceleration by sequences of UV laser filaments with visible and near-infrared pulses}

\author{{Elise Schubert}$^1$, Ali Rastegari$^2$, Chengyong Feng$^2$, Denis Mongin$^1$, Brian Kamer$^2$, {Jérôme Kasparian}$^{1,3}$,  Jean-Pierre Wolf$^1$, Ladan Arissian$^2$, {Jean-Claude Diels}$^2$}
\address{$^1$ Group of Applied Physics, Universit\'e de Gen\`eve, Chemin de Pinchat 22, CH-1211 Geneva 4, Switzerland}
\address{$^2$ University of New Mexico, Center for High Technology Materials, 1313 Goddard SE,
Albuquerque, New Mexico 87106, U.S.A.}
\address{$^3$ Institute for Environmental Sciences, Universit\'e de Gen\`eve,  Boulevard Carl-Vogt 66, CH-1211 Geneva 4, Switzerland}

\ead{jerome.kasparian@unige.ch}
\vspace{10pt}
\begin{indented}
\item[]\today
\end{indented}

\begin{abstract}
We investigate the triggering and guiding of DC high-voltage discharges over a distance of \SI{37}{\cm} by filaments produced by ultraviolet (\SI{266}{\nm})  laser pulses of \SI{200}{\ps} duration. The latter reduce the breakdown electric field by half and allow up to \SI{80}{\%} discharge probability in an electric field of \SI{920}{kV/m}. 
This high efficiency is not further increased by adding nanosecond pulses in the Joule range at \SI{532}{\nm} and \SI{1064}{\nm}. However, the latter statistically increases the guiding length, thereby accelerating the discharge by  a factor of 2. This effect is due both to photodetachment and to the heating of the plasma channel, that increases the efficiency of avalanche ionization and reduces electron attachment and recombination.
\end{abstract}

%
%
%
%
%

\section{Introduction}

Fast switching of high-voltages~\cite{EkberSBGIBSBHL2001,Hendriksa2005}, contact-free connections of moving objects like high-speed trains~\cite{HouarDLAFPMSPC2007}, as well as the control of lightning~\cite{KaspaAAMMPRSSYMSWW2008a}, have motivated intense investigations in the last 25 years. Ultrashort laser filaments~\cite{BraunKLDSM1995,ChinHLLTABKKS2005,couairon2007femtosecond,berge2007ultrashort} are promising candidates in that regard, because they produce ionized channels over extended distances comparable with atmospheric scales~\cite{la1999filamentation,rodriguez2004kilometer,M'ecCADFPTMS2004}. 

Most of the works to date have focused on near-infrared filaments~\cite{ComtoCDGJJKFMMPRVCMPBG2000,LaCCDGJJKMMPRVPCM2000,RodriSWWFAMKRKKSYW2002}, due to the wide availability of the Ti:Sa technology providing \SI{800}{nm} pulses. More recently, excursions further into the IR~\cite{Houard2016}, and even in the mid-infrared~\cite{mongin2016conductivity}, have also shown that the larger filament volume and energy partly balances the low density of free charges related to the less efficient multiphoton ionization for larger wavelengths.


Conversely, the ionization of O$_2$ only needs three photons at \SI{266}{\nm} instead of eight at \SI{800}{\nm}. The multiphoton ionization is therefore more efficient at shorter wavelengths, which results in much higher plasma densities in the case of an ultraviolet laser~\cite{SchwaRD2000,LiuLZLMZ2011}. 
Furthermore, they produce a more homogeneous plasma channel than their near-infrared counterpart~\cite{dergachev2013filamentation}, and can give rise to similar multiple filamentation patterns~\cite{zvorykin2015multiple} that provide numerous filaments in parallel, adding up their conductivities. 
Indeed, the first proposal to guide lightning using ultrashort lasers focused on ultraviolet lasers~\cite{diels1992discharge}. However, experimental demonstrations, that were performed over several tens of centimeters in both pure nitrogen and air, relied on tightly focused ($f$~=~\SIrange[range-phrase=--]{0.3}{1.5}{m}) UV lasers ~\cite{zhao1995femtosecond,  miki1996guiding,rambo2001high} rather than on loosely focused, filamenting beams. 

The lifetime of the plasma is in the range of nanoseconds to tens of nanoseconds for the free electrons~\cite{schubert2016optimal} and in the microsecond range for the ions. The speed of the guided discharge propagation amounts to \SI{e5}{\m \per \s} for a leader regime in gaps or \SIrange[range-phrase=--]{3}{7}{m}~\cite{LaCCDGJJKMMPRVPCM2000}, and \SI{e6}{\m \per \s} in a streamer regime for a shorter gap of \SI{2}{m}~\cite{P'epCVCDJKFMRPCMBLG2001}. Although this propagation speed is 10 times faster than that of unguided discharges, it constitutes a clear limitation to the triggering and guiding of discharges on distances in the meter range and above. Diels and Zhao~\cite{diels1992discharge} suggested to add to the main ionizing pulse either longer pulses in the visible range or a train of ultraviolet pulses, in order to photodetach electrons from both O$_2^-$ and O$^-$ ions, keeping highly mobile free charges, namely electrons, available. Rambo et al.~\cite{rambo2001high} quantified this approach, estimating that at least \SI{5}{\J} at \SI{750}{\nm} in about \SI{10}{\us} are necessary to maintain the plasma density created by \SI{100}{\mJ}, \SI{800}{\fs}, \SI{248.6}{\nm} laser pulses at a repetition rate of \SI{10}{\Hz}, so as to guide discharges over \SI{10}{\m}.

This approach was demonstrated by Méjean et al.~\cite{mejean2006improved}, who observed that adding a 532~nm nanosecond pulse to laser filaments at 800~nm reduced the breakdown voltage by 33~kV/m, i.e. by 5\% as compared to the filaments alone. Ionin et al.~\cite{IoninKLSSSSUZ2012,Ionin2012} showed that long free-running UV pulse sequence following a picosecond UV pulse doubles the breakdown distance. More recently, Zvorykin et al.~\cite{zvorykin2015extended} showed that amplitude-modulated \SI{100}{\ns} pulses at \SI{248}{\nm} obtained by superposing the output of a free-running  cavity discharge to the main pulse reduces the breakdown voltage of laser-triggered discharges. Rambo et al.~\cite{Rambo1999} combined a focused picosecond UV laser with an Alexandrite laser delivering \SI{2}{\micro\s}-long, \SI{210}{mJ} pulses to photodetach electrons. The very long photodetaching pulses are favorable for atmospheric-range scaling, but the limited pulse energy associated to the long pulses drastically limit the corresponding power.

In the present work, we rely on a  \SI{266}{\nm}, \SI{200}{\ps} laser delivering up to \SI{270}{mJ}~\cite{Feng14,Xu14b,Xu14d,Feng17} to generate filaments over \SI{2}{m} and investigate their effect on the triggering and guiding of DC high-voltage discharges over a distance of \SI{37}{\cm} at atmospheric pressure. We show that ultraviolet filaments reduce the breakdown electric field by half. Furthermore, the addition of nanosecond pulses in the Joule range at \SI{532}{\nm} and/or \SI{1064}{\nm} have no measurable effect on the discharge probability, but increases the guiding length, thereby accelerating the discharge by  a factor of  2. This effect is due to both photodetachment and the heating of the plasma channel, that increases the efficiency of avalanche ionization and therefore favors the propagation of a leader along the laser beam path.

\section{Material and methods}

\begin{figure}[t] 
\begin{center}
\includegraphics[width=0.9\columnwidth, keepaspectratio]{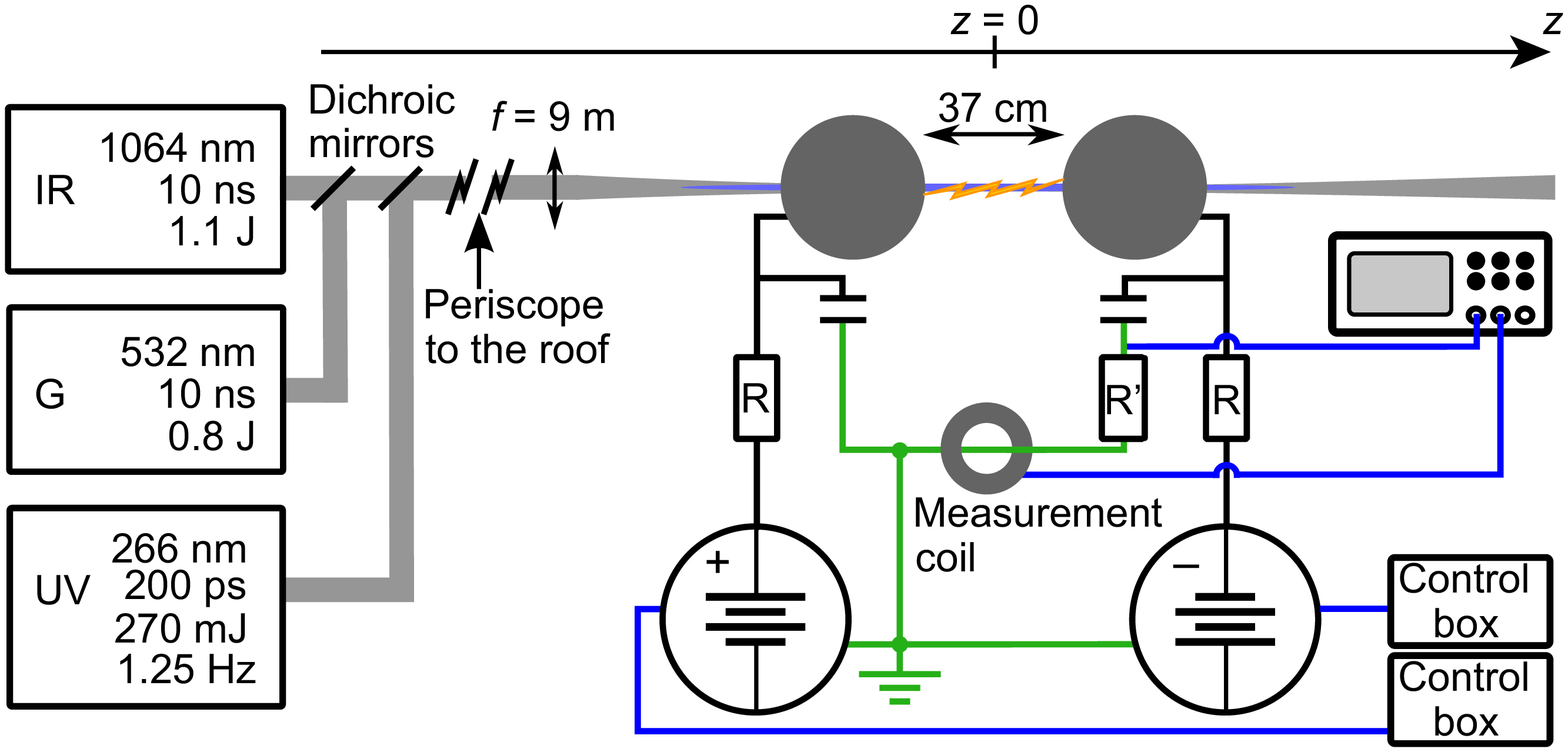}
\end{center}
\caption{Experimental setup. The green (G) and infrared (IR) lasers  are resized by a telescope (not shown) to match the ultraviolet (UV) beam diameter before being coupled with dichroic mirrors.} 
\label{fig:uv_Setup}
\end{figure}

A home-designed ultraviolet laser~\cite{Feng14,Xu14b,Xu14d,Feng17} delivered \SI{200}{\ps} long pulses with \SI{270}{\mJ} output energy at a central wavelength of \SI{266}{\nm} and \SI{1.25}{\Hz} repetition rate. 
Two additional 
Nd:YAG lasers with \SI{10}{\ns} pulse width were used to investigate the effect on the triggering of high-voltage discharges of the heating of 
the plasma channel created by the ultraviolet laser. These secondary lasers respectively delivered \SI{1.1}{\J} at \SI{1064}{\nm} (IR), and \SI{850}{\mJ} at \SI{532}{\nm} (green). The three beam diameters were matched with telescopes and combined so as to ensure that the three beams propagated collinear over \SI{30}{m} to the building roof via a periscope (\Cref{fig:uv_Setup}). The effective ultraviolet energy available between the electrodes was approximately \SI{70}{\%} of the output pulse energy, i.e. \SI{\sim 190}{\mJ}. The delay between the UV and the Nd:YAG lasers was tunable between 0 and \SI{100}{\ns}.
On the roof, the beams were slightly focused by a \SI{9}{\m} focal length fused silica lens into the experimental chamber forming a Faraday cage.
An array of filaments was formed and propagated through the experimental chamber, covering the whole gap between the electrodes.
 
\begin{figure}[t]
\begin{center}
 \includegraphics[width=0.8\columnwidth]{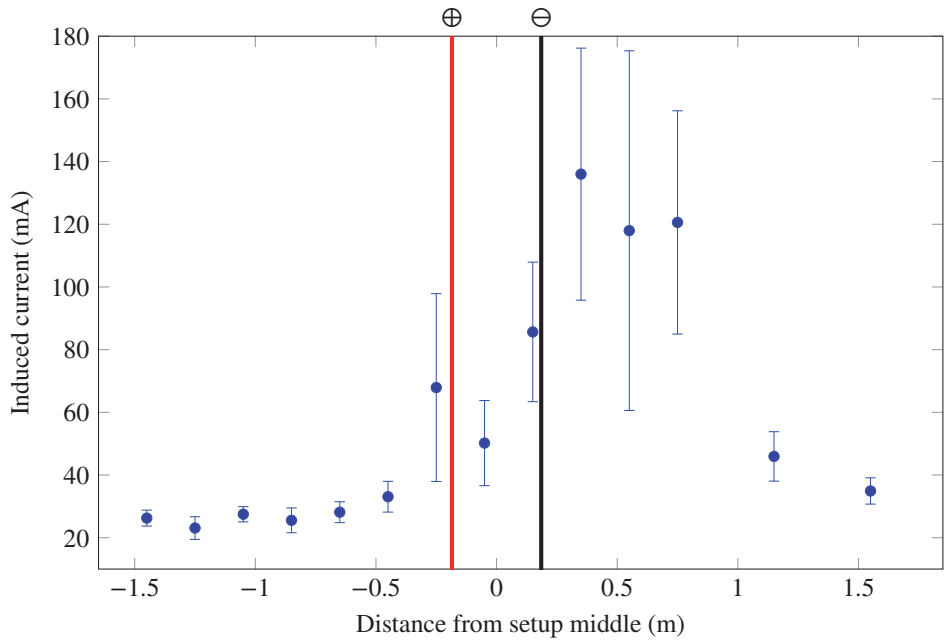}
 \end{center}
 \caption{Transient polarization-induced current by the laser filaments between two square 4~cm$^2$ plane probing electrodes with a 500~kV/m electric field between them. 
 The red and black lines depict the respective positions of the positive and negative electrodes used in the high-voltage experiment.}
 \label{fig:uv_TransverseCurrent}
\end{figure}

We characterized the ultraviolet laser filaments by measuring their conductivity along the beam propagation axis with a capacitive probe~\cite{HeninPKKW2009a,abdollahpour2011measuring}. Two square plane probing electrodes of \SI{4}{\cm\squared} spaced with \SI{2}{\cm} were placed on either side of the beam and moved along the propagation axis. An electric field of \SI{500}{\kV\per\m} was maintained between them. The fast (nanosecond) current due to the laser filament-induced local transient polarization~\cite{abdollahpour2011measuring} was probed through a \SI{10}{\ohm} resistance on the grounding cable. \Cref{fig:uv_TransverseCurrent} displays this current representative of the plasma density as a function of the position along the filaments, that extend over more than \SI{1.5}{m}. 

The two high voltage (HV) hollow spherical electrodes of \SI{40}{\cm} diameter are spaced \SI{37}{\cm}  from one another and  positioned along the beam propagation, in the region where the filaments were the loudest and brightest. Although this position is slightly offset with regard to the measured maximum of ionization, the filament length ensures that the gap between the electrodes is fully covered by the filaments (\Cref{fig:uv_TransverseCurrent}).

Each electrode is charged with the same DC voltage but with opposite polarities by capacitor banks consisting of ten capacitors each, with an equivalent capacitance of \SI{0.2}{\nano\farad} and a resistance of $R = \SI{100}{\mega\ohm}$. The electric circuit was designed such as to minimize the inductance~\cite{feng2016phd}.
This results in a discharge risetime of less than \SI{20}{\ns}. 
In these experiments, the voltage has been varied from \SIrange{\pm 125}{\pm 170}{\kV}, leading to an average electric field from \SIrange{675}{920}{\kV\per\m} between the electrodes. This voltage was measured on a resistor $R'$~=~\SI{136}{k\ohm} (See \cref{fig:uv_Setup}).

The temperature and relative humidity in the experimental chamber varied from \SIrange{27}{35}{\celsius} and \SIrange{13}{23}{\%}, respectively during the campaign. However, the absolute humidity remained rather stable, at  \SI{5.7\pm0.3}{g\per\m\squared}.
A photodiode behind the last mirror triggered the acquisition on a digital oscilloscope of the discharge current to the ground, detected by an induction coil. This current was used to measure the delay between the laser and the breakdown. 
Up to 4 series of \SIrange{150}{250}{ laser shots} were recorded for each laser and voltage configuration, resulting in \SIrange{10}{426}{ breakdown events}. The discharge probability was calculated for each series, and the standard deviation between these series was used to estimate error bars. 

Finally, a camera placed on the side of the beam recorded images of every third laser shot and allowed to characterize the discharge guiding by the laser filaments.

\section{Results and discussion}

\begin{figure}[t]
\begin{center}
 \includegraphics[width=0.8\columnwidth]{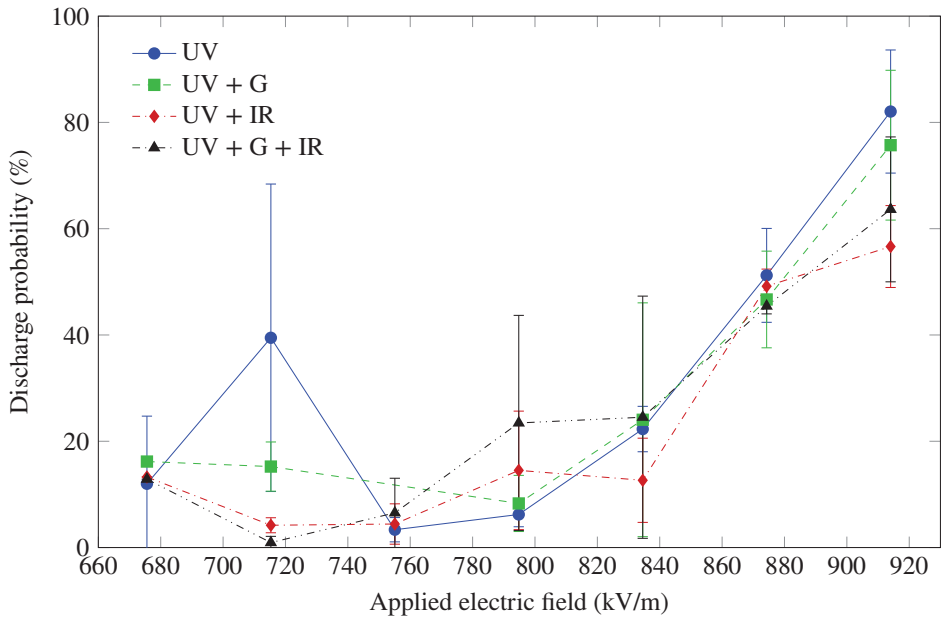}
 \end{center}
 \caption{Discharge probability with respect to the applied electric field for all delays~\SI{\le 50}{\ns} between the ultraviolet (UV) and green (G) and/or infrared (IR) laser pulses. The error bars are the standard deviation between the successive series recorded at each value of the electric field.}
 \label{fig:uv_probaBoum}
\end{figure}

At electric fields up to 920~kV/m investigated during this experiment, no discharge occurs without the laser. All the observed discharges are therefore laser-triggered. 
As displayed on \Cref{fig:uv_probaBoum}, for the ultraviolet filaments alone, the discharge probability within \SI{10}{ns} after the laser pulse increases with increasing electric fields, reaching  \SI{50}{\%} at about \SI{875}{\kV \per \m} and \SI{80}{\%} at \SI{920}{kV/m}. This corresponds to a reduction of more than half as compared with natural breakdown. 
Such reduction is higher than the typical \SI{30}{\%} observed in the case of near-IR filaments~\cite{ComtoCDGJJKFMMPRVCMPBG2000,LaCCDGJJKMMPRVPCM2000,RodriSWWFAMKRKKSYW2002,mejean2006improved}, despite a DC HV supply, that is more favorable to screening.
Therefore, UV filaments are as efficient as tightly focused pulses in the same spectral range~\cite{miki1996guiding,rambo2001high}. 

\begin{figure}[t]
\includegraphics[width=\columnwidth]{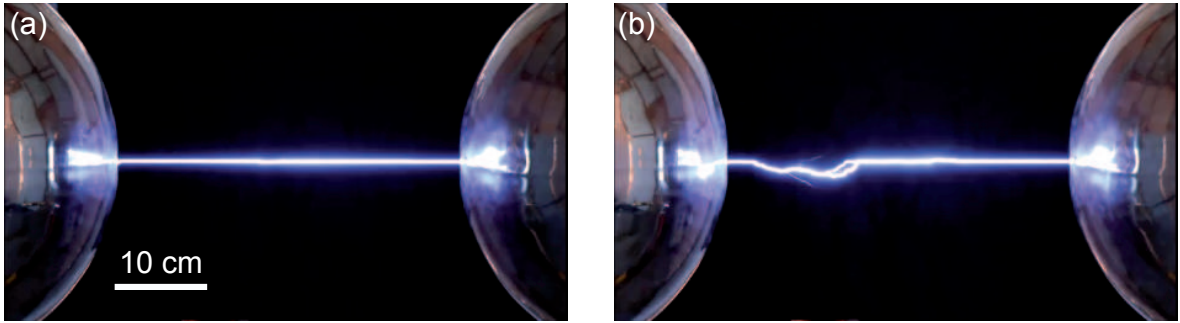}
\caption{Laser-triggered electrical discharges under an electric field of \SI{920}{\kV \per \m}: Fully guided (a);  partly guided (b)}
\label{fig:uv_pictureBoum}
\end{figure}

\begin{figure}[t]
\centering
\includegraphics[width=0.9\columnwidth]{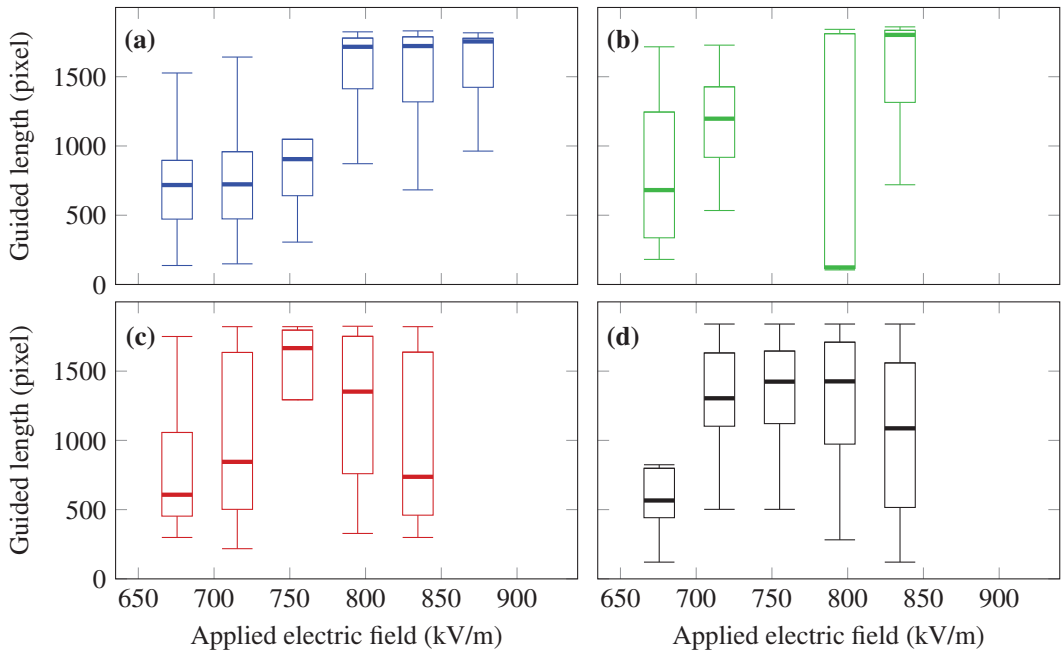}
\caption{Distribution of the guided length of electrical discharges for the UV only (a) and the added \SI{532}{\nm} (b), \SI{1064}{\nm} (c) and both nanosecond lasers (d). Boxes enclose the two central quartiles, and the bold line in the middle marks the median value. Wiskers encompass all data points within 1.5 times the central box width.}
\label{fig:uv_guidedLength}
\end{figure}

 As illustrated in \Cref{fig:uv_pictureBoum}, the laser-triggered discharges give rise to various guiding behaviors, from a full guiding (\Cref{fig:uv_pictureBoum}a) to no guiding at all, through partial guiding (\Cref{fig:uv_pictureBoum}b). 
We calculated the guided length of the discharges on the single-pulse pictures as the number of white pixels along the beam propagation axis between the two electrodes. As displayed on~\Cref{fig:uv_guidedLength}a in the case of the UV laser filaments alone, the guided length tends to increase together with the discharge probability when the electric field increases. A transition occurs for an electric field around \SI{775}{kV/m}. Below this threshold, a majority of discharges are guided over one third to one half of the electrode gap. Above it, most discharges are fully guided.

\begin{figure}[t]
\centering
\includegraphics[width=0.9\columnwidth]{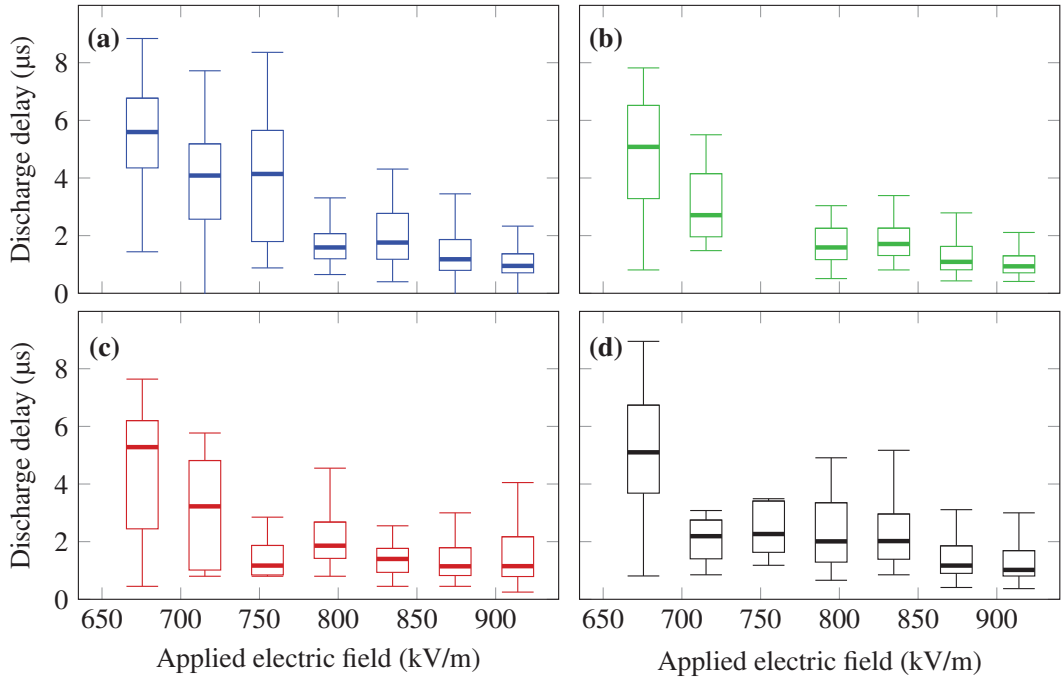}
\caption{Time delay between the laser shot and the triggered discharge with respect to the applied electric field for the UV only (a) and the added \SI{532}{\nm} (b), \SI{1064}{\nm} (c) and both nanosecond lasers (d). 
Boxes enclose the two central quartiles, and the bold line in the middle marks the median value. Wiskers encompass all data points within 1.5 times the central box width.}
\label{fig:uv_LaserBoumDelay}
\end{figure}

Simultaneously, the median delay between the laser pulse and the discharge (\Cref{fig:uv_LaserBoumDelay}a) typically drops by a factor of 2, from 4--6 to \SI{2}{\us}.
The discharge delay includes the time required to both initiate the electron avalanche and propagate the discharge from one electrode to the other~\cite{rambo2001high}.
The simultaneous observation of a guiding length increase by $\sim$20~cm and a delay drop by \SIrange{3}{4}{\us} is consistent with a ten times faster propagation of guided discharges as compared to unguided ones. Furthermore, the corresponding speed increase from \SI{7e4}{m/s} to \SI{2e5}{m/s} points to a streamer-driven discharge regime~\cite{P'epCVCDJKFMRPCMBLG2001}. 

Adding \SI{532}{\nm} and/or \SI{1064}{\nm} nanosecond pulses carrying \SI{\sim1}{\J} of energy each  to the \SI{266}{\nm} filaments has no significant impact on the discharge triggering probability for the investigated electric fields (\SIrange{670}{920}{\kV\per\m}, \Cref{fig:uv_probaBoum}). Consistent results were obtained for delays between the different laser pulses ranging from 0 to 50~ns. Therefore, the data displayed gather measurements for all delays up to \SI{50}{\ns}.
The fact that the nanosecond pulses do not further reduce the breakdown voltage may seem to contradict previous results where they improved the effect of near-infrared filaments~\cite{mejean2006improved}. This discrepancy may be related to the higher reduction of the breakdown threshold caused by the UV filaments alone, as compared to the near-IR filaments (\SI{50}{\%} as compared to \SI{30}{\%}, see above). Such higher efficiency could hide the contributions of the subsequent long pulses.

While they do not significantly influence the discharge probability, nanosecond infrared and visible lasers influence the UV-triggered breakdown process. Their addition to the main UV pulse clearly decreases the electric field threshold where we observe the transition to shorter discharge delays (\Cref{fig:uv_LaserBoumDelay}b--c) and longer guiding lengths (\Cref{fig:uv_guidedLength}b--c). However, this effect is less clear for the guiding length due to larger fluctuations for the strongest electric fields. The contributions of visible and near-infrared pulses add up when the three beams are sent together. In this case the threshold electric field is decreased by about 10\%, at $\sim$700~kV/m (\Cref{fig:uv_guidedLength}d and \Cref{fig:uv_LaserBoumDelay}d).

In order to better understand the effect of the YAG lasers, we numerically investigated the plasma dynamics. As described in detail earlier~\cite{schubert2016optimal}, the model implements multiphoton and tunnel ionization via a Keldysh-PPT approach, avalanche ionization, attachment of electrons to neutral molecules, electron-ion and ion-ion recombination. It also considers the thermal balance via inverse Bremsstrahlung and the energy exchanges between the electrons and heavy species, including the excitation vibrational modes. The model disregards spatial inhomogeneities as well as transport. 

The temperatures and electron densities provided by the model critically depend on the UV intensity in the filaments (e.g. $N_\textrm{e} = \SIrange[range-phrase=-]{1.5}{4.2e22}{m^{-3}}$ and $T = \SIrange[range-phrase=-]{580}{1365}{K}$ for $I = \SIrange[range-phrase=-]{0.5}{0.7}{TW\per\cm\squared}$), that could not be determined precisely in the experiments. Therefore we discuss the effect of the additional YAG laser pulses in terms of trends and ratio.

The main effect of the addition of the YAG lasers is to heat the plasma by inverse Bremsstrahlung.  The green and infrared pulses typically increase the post-pulse electron temperature by \SI{100}{K} and \SI{400}{K}, respectively, which implies losses of \SI{3}{\%} and \SI{10}{\%} of their incident intensity. These contributions add up linearly, so that both pulses together increase the electron temperature by \SI{500}{K}. This difference is preserved over at least a few microseconds during the plasma evolution, i.e., over the full development time of the guided discharge. 

The YAG lasers have virtually no effect on the initial ionization, that mainly occurs via multiphoton ionization. They increase the peak electron density by less than 4\%. However, by heating the plasma, they modify the balance between the avalanche ionization, the electron-ion recombination, and electron attachment to neutrals. 
The avalanche rate increases by almost \SI{10}{\%}, and its effect lasts much longer, up to the YAG pulse duration of \SI{10}{ns}, when both \SI{532}{\nm} and \SI{1064}{\nm} beams are added to the \SI{266}{\nm} beam. Simultaneously, the rates of electron-ion recombination and attachment typically decrease by \SI{25}{\%} and \SI{15}{\%}, respectively. 

These results are consistent with the observation that the electron density produced by a 800 nm filament was only slightly increased by a Nd:YAG pulse sent within 10 ns after the 800 nm pulse and decayed within nanoseconds after that second pulse~\cite{papeer2014extended}.

Besides these thermal effects, both the green and IR YAG pulses efficiently photodetach ions from O$_2^-$ ions, so that during these pulses the net attachment rate typically drops by a factor of 10. As a consequence of the thermal and non-thermal effects of the additional YAG pulses, the free electron density stays almost twice as long (\SIrange[range-phrase=--]{50}{55}{ns} instead of \SIrange[range-phrase=--]{30}{35}{ns}) over its value in natural leaders, namely \SI{1.6e20}{m^{-3}}~\cite{ihaddadene2015increase}.

While the longer free electron lifetime could marginally help the discharge triggering and guiding and reduce the required electric field, we expect that the main effect stems from the hotter channel left behind by the filaments, that reduces the molecular density of the gas. According to Paschen's law~\cite{husain1982analysis,tirumala_analytical_2010} such depleted channel offers a preferential pathway for the discharge~\cite{vidal2000modeling,Houard2016}. It may therefore explain our observation of a better guiding and a faster discharge in the presence of heating YAG pulses. This is  especially true at  \SI{1064}{\nm}, that provides a more efficient heating due to the $1/\omega^2$ dependence of the inverse Bremsstrahlung rate. 

\section{Conclusion}

We demonstrated the high efficiency of UV laser filaments, that are able to reduce the breakdown electric field by half  and allow up to \SI{80}{\%} discharge probability in an electric field of \SI{920}{kV/m}. 
This discharge-triggering efficiency is as high as that of tightly focused UV laser pulses, and higher to that of near-infrared filaments. The addition of further nanosecond pulses in the Joule-range does not improve it further.

Consequently, adding visible or infrared nanosecond pulses to photodetach electrons and heat up the ultraviolet filaments induced electron plasma do not affect the discharge probability. However, it increases the discharge speed and guiding length.
More specifically, we evidence two regimes in the electrical breakdown. At high electric fields, most of the laser-triggered discharges are fully guided, and the laser-to-discharge delay is close to \SI{2}{\us}, corresponding to a connection speed of \SI{2e5}{m/s}. At lower fields, discharges are guided over typically one third to half of their length, and the discharge delay of \SI{5}{\us} corresponds to an average speed of \SI{7e4}{m/s}. Adding green and/or infrared nanosecond pulses to the ultraviolet filaments reduces the critical electric field between these two regimes by \SI{10}{\%} as compared to the ultraviolet filaments alone.

\textbf{Acknowledgements}. We acknowledge financial support from the ERC advanced grant « Filatmo », the ARO MURI grant W911NF1110297, and the DOE award DE-SC0011446.
Technical support by M. Moret was highly appreciated.


\providecommand{\newblock}{}

\end{document}